%% file: main.tex
\documentclass[journal,onecolumn,draftclsnofoot,12pt]{IEEEtran}
\input src/header.tex

\begin{document}
	
	\title{Deep Multi-Emitter Spectrum Occupancy Mapping that is Robust to the Number of Sensors, Noise and Threshold}
	
	\author{Abbas~Termos 
		and~Bertrand~Hochwald,~\IEEEmembership{Fellow,~IEEE} 
		\thanks{Manuscript received: \today. The authors gratefully acknowledge joint funding by the National Science Foundation SpectrumX Center under AST-2132700, and Intel under ECCS-2002921.}
		\thanks{A.\ Termos was with the Department of Electrical Engineering, University of Notre Dame, Notre Dame, IN, 46556 USA (e-mail: atermos@alumni.nd.edu).  He is currently with Qualcomm Technologies Inc.\ in San Diego.} 
		\thanks{B.\ Hochwald is with the Department of Electrical Engineering, University of Notre Dame, Notre Dame, IN, 46556 USA (e-mail: bhochwald@nd.edu)}}

\IEEEtitleabstractindextext{%
	\begin{abstract}
		One of the primary goals in spectrum occupancy mapping is to create a system that is robust to assumptions about the number of sensors, occupancy threshold (in dBm), sensor noise, number of emitters and the propagation environment.  We show that such a system may be designed with neural networks using a process of aggregation to allow a variable number of sensors during training and testing. This process transforms the variable number of measurements into approximate log-likelihood ratios (LLRs), which are fed as a fixed-resolution image into a neural network.  The use of LLR's provides robustness to the effects of noise and occupancy threshold.  In other words, a system may be trained for a nominal number of sensors, threshold and noise levels, and still operate well at various other levels without retraining.  Our system operates without knowledge of the number of emitters and does not explicitly attempt to estimate their number or power.  Receiver operating curves with realistic propagation environments using topographic maps with commercial network design tools show how performance of the neural network varies with the environment.  The use of very low-resolution sensors in this system can still yield good performance.
	\end{abstract}  
	
	\begin{IEEEkeywords}
		Convolutional neural network, supervised learning, spectrum occupancy mapping, distributed sensing, robustness
\end{IEEEkeywords}}

\maketitle

\IEEEdisplaynontitleabstractindextext

\IEEEpeerreviewmaketitle

\section{Introduction}

Spectrum occupancy mapping is the process of identifying occupied frequency bands using power measurements from distributed sensors, where the occupancy maps are decisions of whether power (in dBm) exceeds a threshold in various locations over an extended region, not just where the sensors are. 
In this work, we form these maps without estimating signal power or emitter locations directly, with the goal of creating a system that is robust to assumptions about the propagation environment, number of sensors and emitters, occupancy threshold, and sensor noise. 

\subsection{Background}
\label{subsec:background}
Classical methods for estimating power maps from which occupancy can be derived often postulate models that describe the sensor measurements from a set of emitters. These methods are largely parametric, and attempt to solve the inverse problem of finding the emitters from the measurements. The solutions are developed under simplified path loss relationships between sensors and a single emitter, such as free-space \cite{li2003energy}, or log-normal shadowing \cite{liu2006analysis,lin_accurate_2013}.  Multiple emitters in free-space are handled using a maximum-likelihood-based approach \cite{sheng2005} provided that the number of emitters is known. When the number of emitters is not known, compressive sensing approaches are used \cite{feng2009multiple,bazerque2010}, where the emitter locations and powers are found simultaneously. However, these methods are not robust to emitter numbers, or locations, which are assumed to be sparse, and perform poorly in complex propagation environment due to violations of their recoverability conditions.

On the other hand, non-parametric methods use interpolation techniques in which the power level at any location is represented as a linear combination of the measurements. Examples of such methods include inverse-distance weighting \cite{shepard1968two,shawel2018deep}, K-nearest neighbors (K-NN) \cite{cover1967nearest}, radial basis function interpolation \cite{bishop2006pattern} and Kriging \cite{chakraborty2017specsense}. For a detailed discussion of classical methods, see \cite{bishop2006pattern}. Such methods are generally sensitive to emitter powers, locations, number, and the propagation environment, as they rely on statistical, not physical, relationships. The appeal of such classical methods is that they do not need to be trained and work well when the model matches the real environment.

When an accurate model cannot be prescribed, neural networks can address the problem of determining power maps from sensor measurements.  Deep neural networks have been demonstrated to learn power maps constrained to a single unknown emitter when sensor data is acquired using a parametric forward model \cite{hashimoto2020sicnn}, ray-tracing \cite{krijestorac2021spatial}, or measurement campaigns \cite{zhang2021missing}. Many of these early efforts look at only a small number of emitters in the same band.
Training and testing are then often done with a single emitter \cite{han2020power}, or less than three emitters \cite{teganya2020data,teganya2020deep}. A larger number of emitters is tackled in \cite{shrestha2021deep,shrestha2022deep} by assuming that the power spectra of the emitters do not completely overlap in frequency. The neural network task is then reduced to computing power maps for a small number of emitters per frequency band, and requires estimation of the number of emitters.  We compare our approach with both neural network and non-neural network alternatives in Section \ref{sec:robustness}. 

\begin{figure}[!t]
	\centering
	\includegraphics[width = 0.5\textwidth]{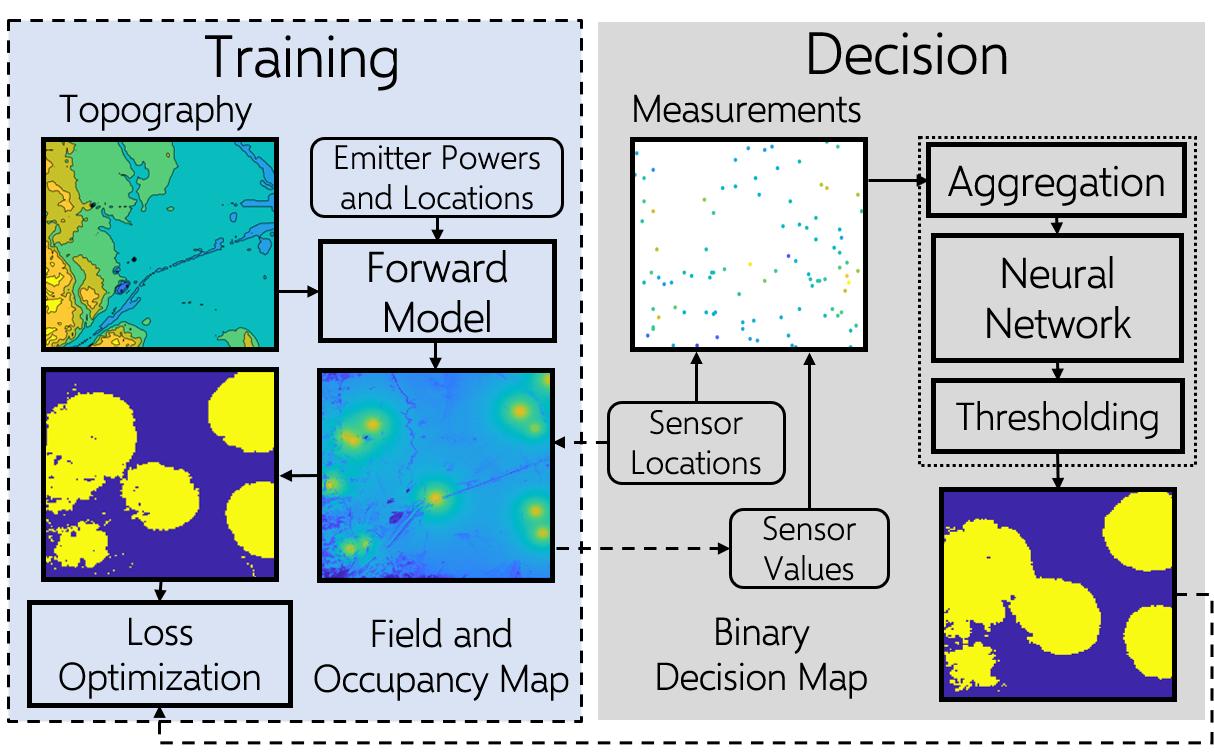}
	\caption{\footnotesize A neural network-based system for producing decision maps from sensor measurements. 
		The portion on the right aggregates sensor measurements into fixed-resolution input images and produces decision maps, also in the form of an image, using a neural network.  The portion on the left supervises the training with sensor measurements and their corresponding occupancy maps for optimizing the loss function. The dashed traces indicate information flow during training.
	}
	\label{fig:mlblock}
\end{figure}

\subsection{Contributions}
%
The system shown in Fig.~\ref{fig:mlblock} is used to determine spectrum occupancy maps and has the following features:
\subsubsection{\bf Ability to handle arbitrary propagation environment and number and locations of emitters} 
We present the problem of occupancy mapping as a binary decision problem for the neural network: given a location and occupancy threshold, we determine if power at a given frequency exceeds the threshold, without knowledge of the number of emitters or an explicit step to estimate their number.  Thus sensitivity to ``model order" is eliminated.  Variable and unlimited numbers of emitters are allowed during training and testing, all occupying the same frequency band.

\subsubsection{\bf Robustness} 
We present a simple and novel log-likelihood (LLR) approach to calculate the inputs to the neural network from sensors measurements. We show that this naturally addresses what to do with locations that have no sensors, and provides robustness to the occupancy threshold, number of sensors and noise.  A variable number of sensors is allowed, eliminating the constraint that the number of inputs to the neural network must be prescribed.

Using a figure of merit, we demonstrate how the performance of the neural network varies with the environment by computing receiver operating curves with realistic propagation environments using topographic maps with commercial network design tools.  We introduce the usage of a ``threshold-to-noise ratio" to replace signal-to-noise ratio as a figure of merit.  We also show that, perhaps surprisingly, very low-resolution sensors can still be used to get good performance.

The paper is organized as follows. Section \ref{sec:problem} introduces the setup, and problem formulation. Section \ref{sec:nn} describes the details of the LLR-based input aggregation process, the network architecture, the dataset and establishes a baseline performance for robustness studies. Section \ref{sec:robustness} demonstrates the robustness of the decision system to the occupancy threshold, varying the number of sensors, and additive noise. Section \ref{sec:tnrOnebit} presents some practical sensor considerations with regard to threshold-to-noise ratio and one-bit measurements.

\section{Problem Formulation}
\label{sec:problem}
\subsection{System setup}
Consider a region $\region \subset \mathbb{R}^2$. Denote the set of emitter (coordinate) locations by $\emitterSet = \left\{\emitterLoc_{l} \right\}_{l=1}^{\nEmitters}$, where $\nEmitters$ is the number of emitters, and $\emitterLoc_l \in \region$ $\forall l$. Denote the set of emitter powers by $\emitterPowersSet = \left\{\powEl_l \right\}_{l=1}^{\nEmitters}$, where $\powEl_l \in \mathbb{R}$ and $\powEl_l>0$ $\forall l$, expressed in units of Watts; these quantities are generally unknown and random. A sensor at a location $\fieldLoc \in \region$ then measures the power
\begin{equation}
	\field(\fieldLoc) \equiv \field(\fieldLoc,\nEmitters,\emitterLoc_{1},\emitterLoc_{2},\cdots,\emitterLoc_{\nEmitters}, \powEl_{1},\powEl_{2},\cdots,\powEl_{\nEmitters}),
	\label{eq:field}
\end{equation} 
where the right-hand side of \eqref{eq:field} shows the explicit dependence of the measurements on the emitters. We refer to $\field(\fieldLoc)$ as the power field, or simply, field.

Denote the set of sensor locations by $\sensorSet = \{ \sensorLoc_j \}_{j=1}^{\nSensors}$, where $\nSensors$ is the number of sensors and $\sensorLoc_j \in \region$ $\forall j$; these quantities are generally known, but not under our control. It is assumed that the sensors are all tuned to the same center frequency and bandwidth of interest. It is also assumed that the emitters are all active in this band. While this frequency is an important parameter in determining the nature of propagation of energy throughout the region, our methodology is not constrained to any given band or propagation model.  Hence, we omit mention of frequency until Section \ref{sec:occ-map}, where simulations are shown for a given frequency.

We define the partition $\gridSet = \{\gridSet_k\}_{k=1}^{\nGrid}$, with sub-regions $\gridSet_k \subset \region$, as a collection of non-overlapping sets that cover $\region$, where $\nGrid$ is the number of sub-regions. We define the occupancy in a sub-region $\nntarget(\gridSet_k)$ as the indicator function for whether the field, averaged over $\gridSet_k$, exceeds a predetermined occupancy threshold $\occThresh$, or 
\begin{equation}
	\nntarget(\gridSet_k) = \mathbbm{1}\{\meanField \ge \occThresh\},
	\label{eq:nntarget}
\end{equation}
where $\meanField$ is the mean field over $\gridSet_k$, computed over all $\fieldLoc \in \gridSet_k$, and $\mathbbm{1}\{\cdot\}$ is an indicator function that is one if its argument is true and zero otherwise. Thus, if the average $\field(\fieldLoc)$ exceeds $\occThresh$ in $\gridSet_k$, then $\nntarget(\gridSet_k)=1$.  Otherwise, it is zero. Typically, $\gridSet$ rasterizes $\region$ along a regular grid and the occupancy maps are images of fixed resolution. The occupancy map is then defined as $\nntarget(\gridSet) \equiv \left\{\nntarget(\gridSet_k) \right\}_{k=1}^{\nGrid}$.  This definition of occupancy naturally induces two hypotheses for every sub-region $\gridSet_k$, namely, $\mathcal{H}_{0,k}: \meanField < \occThresh$ where $\nntarget(\gridSet_{k}) = 0$, and $\mathcal{H}_{1,k}: \meanField \ge \occThresh $ where $\nntarget(\gridSet_{k}) = 1$.

The choice of $\nGrid$ determines the resolution of the occupancy map, and hence will generally be a function of the intended use of the map -- small $\nGrid$ gives coarse occupancy resolution, while large $\nGrid$ gives refined occupancy measurements. In our results, $\nGrid$ is chosen such that the variation of the field within a sub-region $\gridSet_k$ is less than $2 \,\rm{dB}$ on average.

Although the sensor locations are known, their position and number are random and not under our control. The sub-regions form a grid that covers the region, and the number of sub-regions that include at least one sensor is smaller (by orders of magnitude) than the total number of sub-regions for which occupancy decisions are desired. 

The number, location, and power of the emitters at the frequency or band of interest are entirely unknown.  Hence, the ground truth of which sub-regions are occupied varies over the region with the number and location of the emitters, and the neural network has the task of mapping the random sensor measurements to a decision map, denoted by $\nndecision(\gridSet)$, throughout the region.


Realistic forward modeling software and topographical maps are used to generate $\field(\fieldLoc)$ and ensure that the generated occupancy maps are accurate. No forward model is assumed by the network, except what was indirectly learned during training. No explicit steps of estimating the number of emitters or their locations are performed. A sequence of measurements produces a sequence of decision maps, but temporal prediction and modeling are not considered. 

\subsection{Problem statement} 
Given a set of sensor locations $\sensorSet \subset \region$, and measurements $\meas(\sensorLoc), \, \sensorLoc \in \sensorSet$, we wish to use the neural network system shown in Fig.~\ref{fig:mlblock} to determine occupancy decisions $\nndecision(\gridSet_k)$ over all $\gridSet_k \in \gridSet$ covering $\region$. The sensors measure power, and we consider an aggressively small number of sensors, i.e. $\nSensors \ll \nGrid$. The neural network-based decision system produces a decision map in two steps:
\begin{enumerate}
	\item The sensor measurements $\meas(\sensorLoc)$ are aggregated into an image whose resolution is determined by $\gridSet$
	\item A neural network forms a binary decision map over all of $\gridSet$ from this image
\end{enumerate}
The neural network-based decision system produces a decision map over $\gridSet$, whether or not a sensor occupies every $\gridSet_k$

\begin{figure*}[!t]
	\centering
	\subfloat[]{
		\centering
		\includegraphics[trim={2.3cm 1.0cm 1.3cm 0.6cm},clip,width=0.16\textwidth]{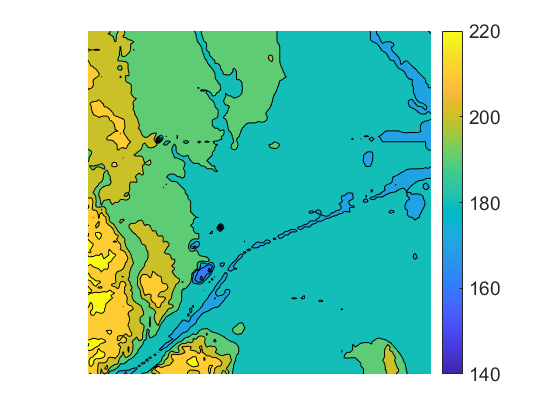}
		\label{fig:alts-chicago}}
	\subfloat[]{
		\centering
		\includegraphics[trim={2.3cm 1.0cm 1.3cm 0.6cm},clip,width=0.16\textwidth]{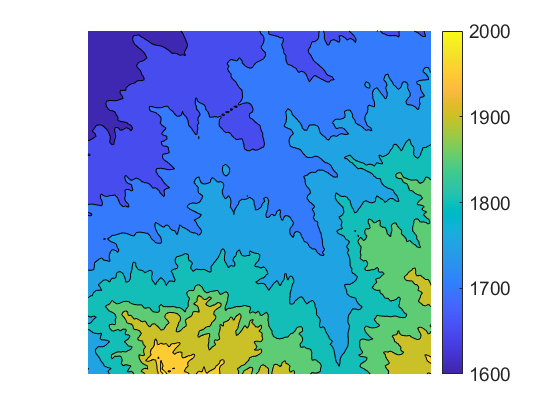}
		\label{fig:alts-denver}}    
	\subfloat[]{
		\centering
		\includegraphics[trim={2.3cm 1.0cm 1.3cm 0.6cm},clip,width=0.16\textwidth]{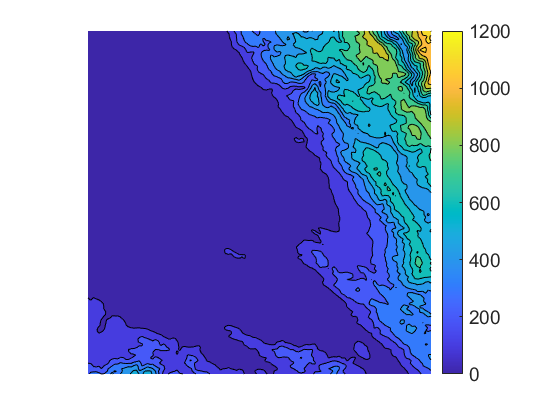}
		\label{fig:alts-sanjose}}    
	\subfloat[]{
		\centering
		\includegraphics[trim={3.1cm 1.3cm 1.0cm 0.7cm},clip,width=0.16\textwidth]{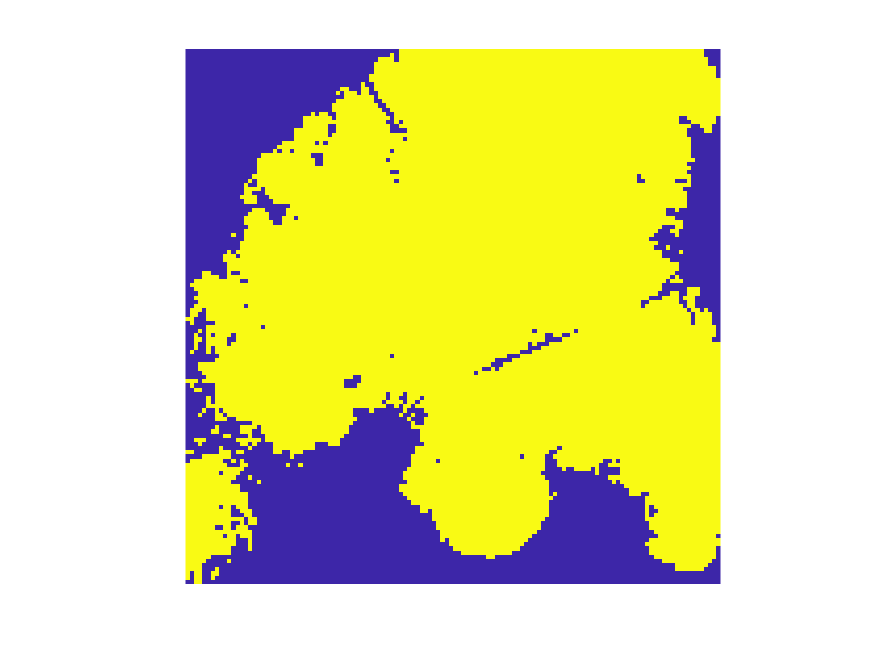}
		\label{fig:occ-chicago}}
	\subfloat[]{
		\centering
		\includegraphics[trim={3.1cm 1.3cm 1.0cm 0.7cm},clip,width=0.16\textwidth]{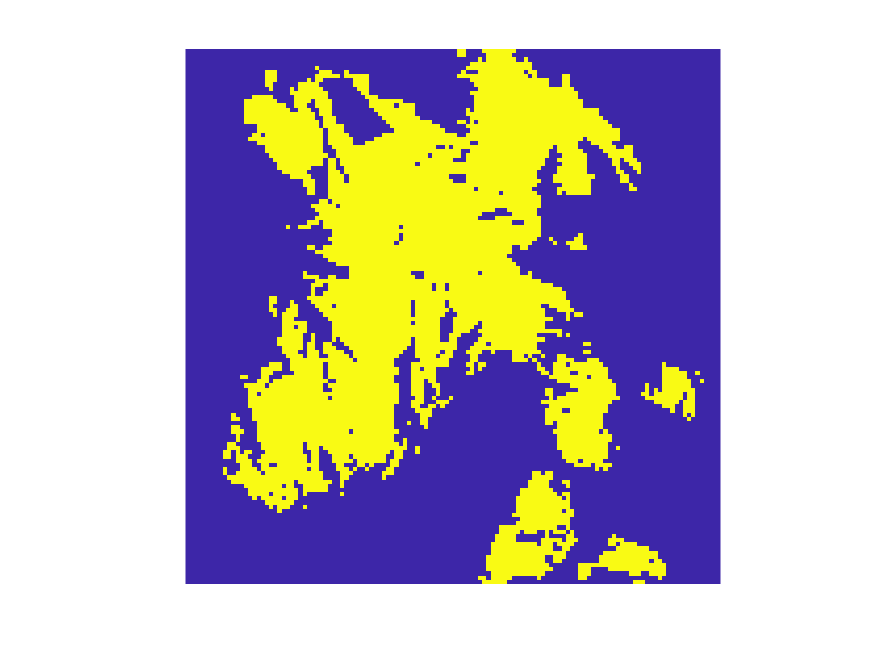}
		\label{fig:occ-denver}}
	\subfloat[]{
		\centering
		\includegraphics[trim={3.1cm 1.3cm 1.0cm 0.7cm},clip,width=0.16\textwidth]{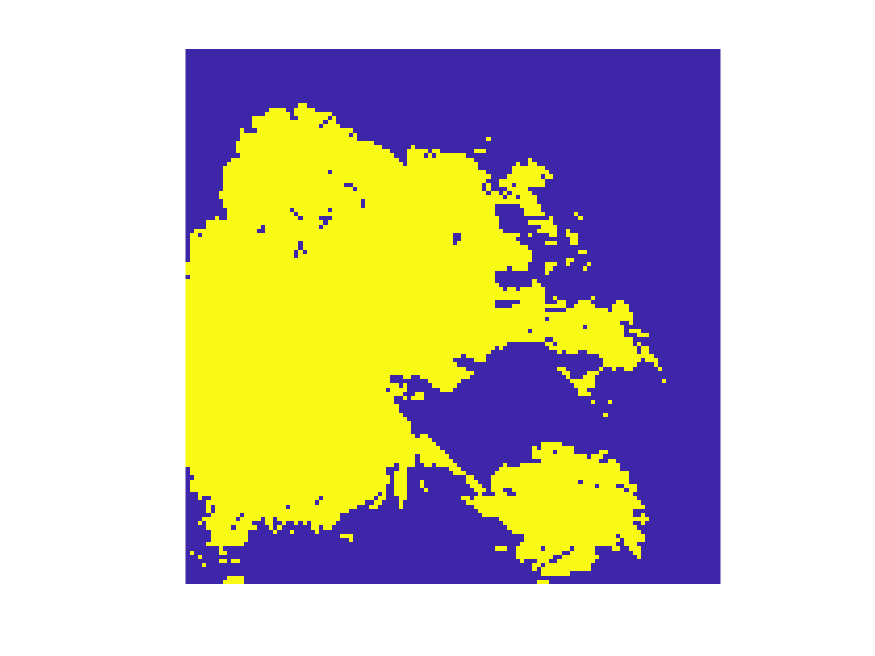}
		\label{fig:occ-sanjose}}

	\subfloat[]{
		\centering
		\includegraphics[trim={2.3cm 1.0cm 1.3cm 0.6cm},clip,width=0.16\textwidth]{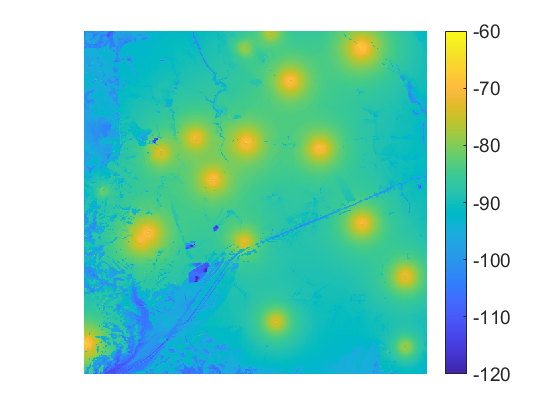}
		\label{fig:field-chicago}}
	\subfloat[]{
		\centering
		\includegraphics[trim={2.3cm 1.0cm 1.3cm 0.6cm},clip,width=0.16\textwidth]{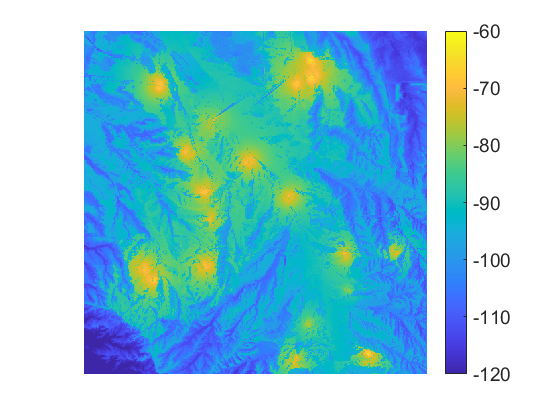}
		\label{fig:field-denver}}
	\subfloat[]{
		\centering
		\includegraphics[trim={2.3cm 1.0cm 1.3cm 0.6cm},clip,width=0.16\textwidth]{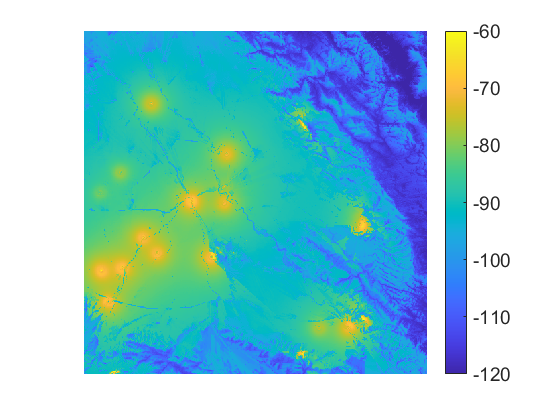}
		\label{fig:field-sanjose}}
	\subfloat[]{
		\centering
		\includegraphics[trim={3.1cm 1.3cm 1.0cm 0.7cm},clip,width=0.16\textwidth]{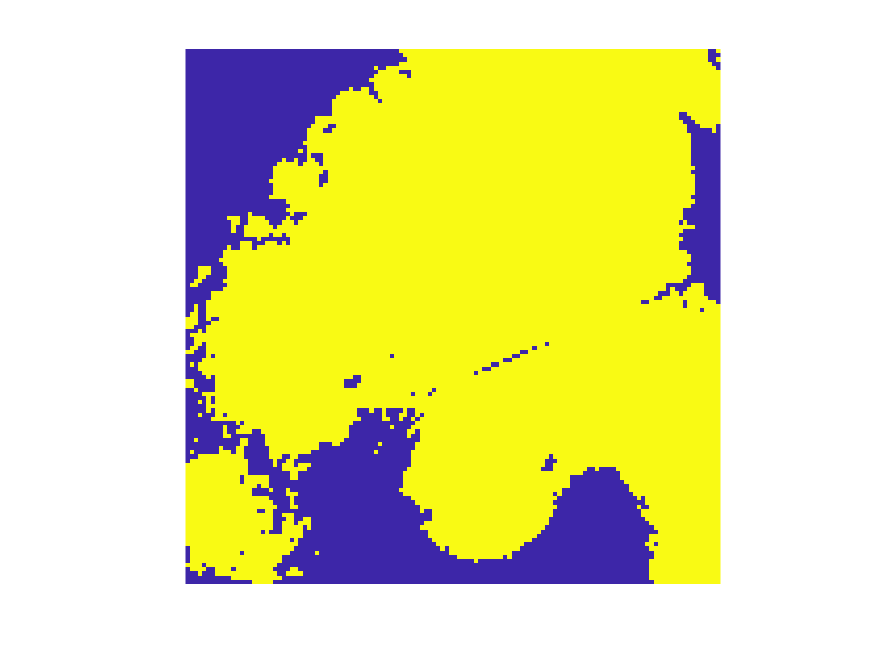}
		\label{fig:pred-chicago}}
	\subfloat[]{
		\centering
		\includegraphics[trim={3.1cm 1.3cm 1.0cm 0.7cm},clip,width=0.16\textwidth]{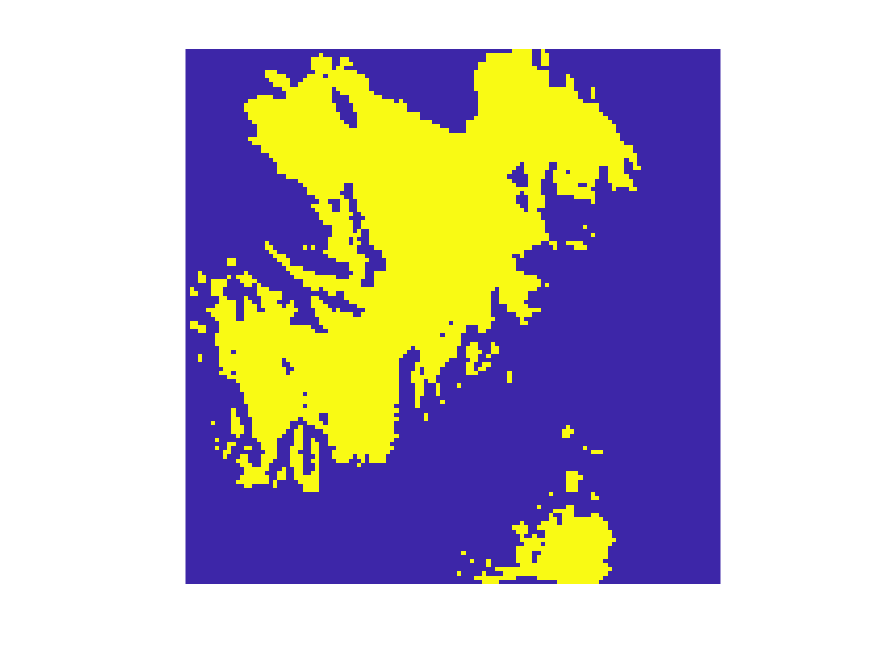}
		\label{fig:pred-denver}}
	\subfloat[]{
		\centering
		\includegraphics[trim={3.1cm 1.3cm 1.0cm 0.7cm},clip,width=0.16\textwidth]{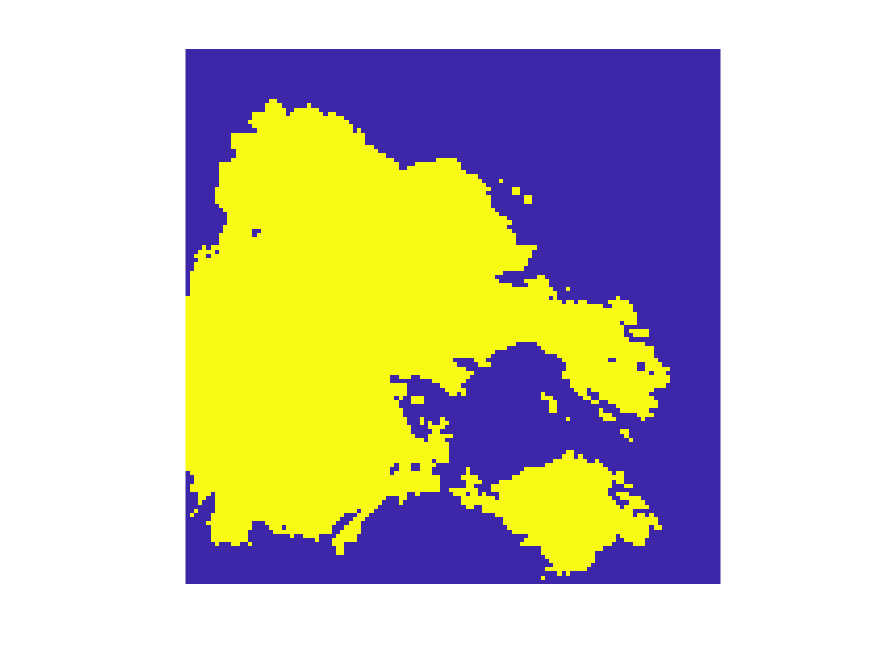}
		\label{fig:pred-sanjose}}
	\caption{\footnotesize Fields \eqref{eq:field}, occupancy maps \eqref{eq:nntarget} and decision maps \eqref{eq:nndecisions} in a variety of region topographies. (a)--(c): contours of the altitude profiles (in meters) for (a) Chicago, (b) Denver and (c) San Jose. (g)--(i): fields (in dBm) with $\nEmitters=18$ corresponding to the regions in (a)--(c). (d)--(f): occupancy maps with $\occThresh=-90\,\rm{dBm}$. Yellow (blue) indicates that $\meanField$ is larger (smaller) than $\occThresh$. (j)--(l): decision maps, where comparing (d) to (j) and (f) to (l) shows that the occupancy maps are accurately estimated in Chicago and San Jose, resulting in a low error rate (about $8\%$). Comparing (e) to (k) shows that the occupancy patterns in Denver are harder to estimate resulting in higher error rate (about $15\%$). In these figures, the regions are $655.36, \, \rm{km}^2$ and $\nGrid=128\times128$.  The details of the training process and performance are described in Section \ref{sec:occ-map}, and the error rate is tabulated in Table \ref{tab:train-test-stats}. }
	\label{fig:field}
\end{figure*}

\section{Neural Network Architecture and Dataset}
\label{sec:nn}

In this section, we present a convolutional neural network (CNN) with an encoder-decoder structure, and show how it is used to produce decision maps. In particular, we describe the measurement model, and the aggregation of sensor measurements as log-likelihood ratios (LLRs) into images of fixed size suitable as inputs to a neural network. Moreoever, we describe the generation of the dataset from occupancy maps and how it is used to train and test the neural network.


\subsection{Regions and occupancy maps}
\label{sec:occ-map}

Occupancy maps used in training are taken as ground truth, and thus the degree to which the decision maps represent reality is determined by the choice of the forward model. The methods we present are not constrained to any particular model, and can work with free-space, dominant path loss \cite{wahl2005dominant}, ray-tracing simulators and learned radio maps such as \cite{saito2019two,imai2019radio,sotiroudis2020deep,ito2021radio,levie2021radiounet}.
Software packages that provide accurate forward models are used extensively by wireless operators when designing their commercial wireless deployments. We utilize Forsk's Atoll \cite{atoll}, which is especially useful for outdoor environments. Atoll provides control over the region topography, antenna patterns, heights, and tilts, emitter powers and locations, and allows field simulations over many emitter configurations.

We select regions $\region$ with distinct topographies in the vicinity of three cities: Chicago, Denver and San Jose as shown in the contour maps Fig.~\ref{fig:alts-chicago}--\ref{fig:alts-sanjose}. Chicago mostly comprises flat (80m variation) low-altitude areas, which allow the emitter power to propagate without significant shadowing. In a flat environment, occupancy is predictably highly correlated with proximity to an emitter.  In contrast, Denver has an altitude profile with a variation of over $500\,\rm{m}$, and occupancy maps have complex patterns due to shadowing by the landscape. San Jose's altitude profile is characterized by a mixture: flatness in the valley, with high altitudes in the surrounding hills and mountains. Fig.~\ref{fig:field-chicago}--\ref{fig:field-sanjose} show corresponding examples of $\field(\fieldLoc)$, and Fig.~\ref{fig:occ-chicago}--\ref{fig:occ-sanjose} show their corresponding occupancy maps $\nntarget(\gridSet)$ for $\nEmitters=18$. To obtain the topography of the region $\region$, we use Esri's ArcGIS to project the digital elevation models, publicly available at the United States Geological Survey (USGS) website \cite{usgs}, into a raster format usable by Atoll. All of the square regions are $655.36 \,\rm{km^2}$. The topography in each region is fed to Atoll as a raster image of size $512 \times 512$, or a resolution of $50 \times 50 \, \rm{m}^2$. The partition $\gridSet$ contains $\nGrid=128 \times 128$ sub-regions, each having an area $200 \times 200 \, \rm{m^2}$. All emitters use 11dBi omni-directional 0-degree tilt antennas at $2100\, \rm{MHz}$ mounted $20\rm{m}$ above ground level. The bandwidth considered is $10 \, \rm{MHz}$.



\begin{figure*}[!t]
	\centering
	\includegraphics[width = \textwidth]{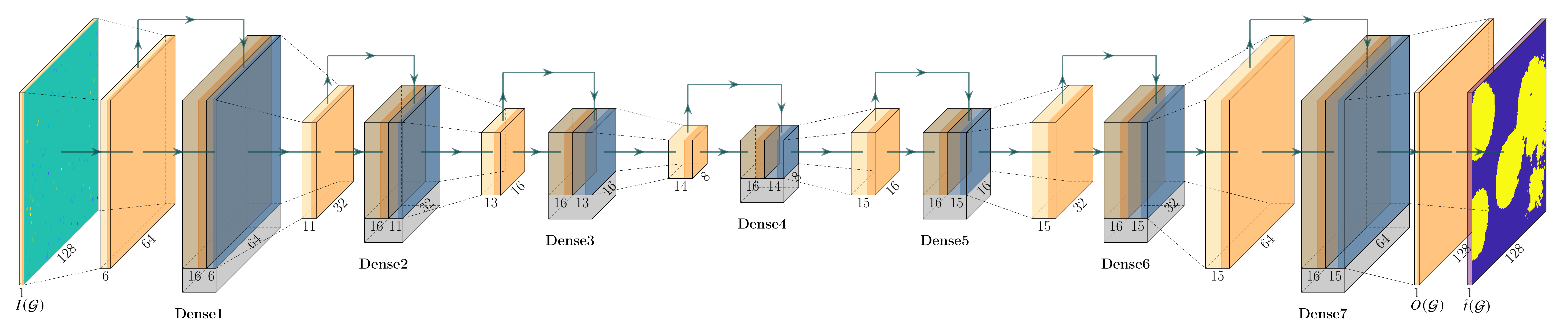}
	\caption{\footnotesize Encoder-Decoder architecture where the (square) feature map dimensions are indicated.  The output of a dense block is the concatenation of the output of a convolution (orange) and the copied input (blue). Convolution blocks associated with transition layers shrink the input size during encoding, and expand during decoding. The output $\nnoutput(\gridSet)$ passes through a Sigmoid layer and a threshold detector to produce the decisions $\nndecision(\gridSet)$.}
	\label{fig:nn-arch}
\end{figure*}

\subsection{Inputs and aggregation}
\label{subsection:agg}


The sensors measure power $\meas(\sensorLoc)$ at a given frequency and bandwidth at locations $\sensorLoc \in \sensorSet$. 
A measurement made in a sub-region $\sensorLoc \in \gridSet_k$ can be modeled locally as
\begin{equation}
	\meas(\sensorLoc) = \meanField  + \modelNoise(\sensorLoc),
	\label{eq:modelNoise}
\end{equation}
where $\modelNoise(\sensorLoc)\sim\mathcal{N}(0,\modelVar)$ is the error of representing the mean field $\meanField$ by a single measurement $\meas(\sensorLoc)$. The error in \eqref{eq:modelNoise} is generally a function of size; the smaller the sub-region, the smaller the error.  In a single-sensor system, an optimal decision of occupancy is obtained by computing the likelihood function $\likelihood\left(\meas(\sensorLoc)\mid \meanField\right)$ using \eqref{eq:modelNoise} to decide between the hypotheses $\mathcal{H}_{0,k}:\,\nntarget(\gridSet_k) = 0$ versus $\mathcal{H}_{1,k}:\,\nntarget(\gridSet_k)=1$. Since $\meanField$ is unknown, a generalized likelihood-ratio test is used:
\begin{align}
	LLR(\sensorLoc) &= \log\dfrac{\max_{H_1:\meanField\ge\occThresh}\likelihood\left(\meas(\sensorLoc)\mid\meanField\right)}{\max_{H_0:\meanField<\occThresh}\likelihood\left(\meas(\sensorLoc)\mid \meanField\right)} \label{eq:llr} \\
	& =
	\dfrac{(\meas(\sensorLoc) - \occThresh)|\meas(\sensorLoc) - \occThresh|}{2\modelVar}.
	\label{eq:llrNoiseless}
\end{align}
Because we have a multi-sensor system, we do not make local decisions but instead feed these LLRs as
inputs to the neural network.  We can assume $\modelVar=1/2$ because constant factors do not affect the training of the network.  Moreover, since the measurements generally have many orders of magnitude dynamic range, we have found that dropping the absolute value term in \eqref{eq:llrNoiseless} aids in numerical stability. Thus, the sensors report the approximate LLRs
\begin{equation}
	\sensdecision(\sensorLoc) = \meas(\sensorLoc)-\occThresh.
	\label{eq:sensSoftDec}
\end{equation}
This approximation is an intuitively pleasing metric representing the difference between the measured sensor value and the threshold.  The larger the metric, the more sure we are that the measurement exceeds the threshold, and conversely.  Where there are no sensors, we set $\sensdecision(\fieldLoc) = 0$ for $\fieldLoc \in \region \backslash \sensorSet$, representing no information of whether the field is above or below the threshold. We show in Section \ref{sec:noise} that this method of LLRs lends itself well to noisy measurements, and results in an expression for $\sensdecision(\sensorLoc)$ that is a generalization of \eqref{eq:sensSoftDec}. 

We then form the input (image) to the neural network as
\begin{equation}
	\nninput(\gridSet_k) = 
	\frac{1}{\llrNormalize}\frac{1}{ |\gridSet_k\cap\sensorSet|}\sum_{\fieldLoc \in \gridSet_k} \sensdecision(\fieldLoc),
	\label{eq:nninput}
\end{equation}  
which pools the measurement of all the sensors in $\gridSet_k$, where $|\gridSet_k\cap\sensorSet|$ is the number of sensors contained in $\gridSet_k$. Thus, the neural network input is the aggregated LLRs in every sub-region that has at least one sensor. For sub-regions containing no sensors, the LLRs are naturally zero. We introduce a normalization constant $\llrNormalize$ that scales the values encountered to be in a range typical for the input to a neural network.  For example, choosing $\llrNormalize$ such that the image $\{\nninput(\gridSet_k)\}_{k=1}^{\nGrid}$ has unit variance works well in practice. 

Because we typically consider a square region, the aggregation process forms an image of size $\sqrt{\nGrid}\times\sqrt{\nGrid}$.  We are interested in constructing a binary decision map over all the sub-regions, including those for which $\nninput(\gridSet_k)=0$. This resembles the problem of image inpainting, where incomplete images with regular \cite{yan2018shift} or irregular \cite{liu2018image} holes are recovered from available pixels. CNNs commonly used in these problems employ an encoder-decoder structure, which allows the translation of one type of (input) image, to another type of (output) image. In the problem we consider, the hole-to-image ratio is $1-\nSensors/\nGrid\approx 0.99$.

\subsection{Neural network}
\label{subsec:nn}

The encoder-decoder structure that computes $\nndecision(\gridSet)$ comprises $22$ convolutional layers. Each convolutional layer results in a number of output ``images'' called feature maps. To avoid accuracy saturation caused by training deep CNNs, we incorporate dense blocks which concatenate output feature maps with input feature maps. Dense blocks are the building blocks of DenseNets \cite{huang2017densely}, which demonstrate excellent performance for classification on the ImageNet dataset \cite{deng2009imagenet}. A dense block is defined in terms of the number of convolutional layers $\denseLayerDepth$ and growth rate $\denseGrowth$. The growth rate $\denseGrowth$ specifies the number of feature maps produced by each layer within a dense block. Since a dense block operates by concatenation, the input and output feature sizes must match.  If there are $\denseGrowth_0$ feature maps at the input of a dense block with $\denseLayerDepth$ layers, then the number of feature maps at the output of the dense block is $\denseGrowth_0 + \denseGrowth \times \ell$.  To allow the CNN to process feature maps at different resolutions, dense blocks are followed by transition blocks that either downsample or upsample feature maps depending on the encoding or decoding operation.

The network structure is shown in Fig.~\ref{fig:nn-arch}. The input to the network $\nninput(\gridSet)$ consists of the aggregated LLRs using \eqref{eq:nninput}. The first yellow block is the output feature maps of a 2D k21s2p10 ($\rm{kernel\, size}=21$, $\rm{stride}=2$, $\rm{padding}=10$) convolution layer with six output channels. These initial features are the input to the encoder. We refer to a convolution block as a combination of three layers: batch normalization (BatchNorm), Rectified-linear unit (ReLU) and convolution layers. The encoder is a dense CNN with three dense blocks having $\denseLayerDepth=1$ and $\denseGrowth=16$, and three down-sampling transition blocks. Within a dense block, the convolution layers are k3s1p1. A Transition block consist of two convolution blocks. The first uses a k1s1p0 convolution layer which learns pooling across feature maps, and halves the number input channels. The second uses a k3s2p1 convolution layer, which halves the size of the input feature maps. Therefore the transition blocks at the encoder have outputs whose dimensions are half those of their inputs. The decoder is also a dense CNN with four dense blocks, and three up-sampling transition blocks. The decoder dense blocks are identical to the encoder dense blocks. Transition blocks at the decoder differ from those at the encoder in the second convolution blocks which use convolution transpose layers of k3s2p1 to upsample the feature maps by a factor of two. The final layer of the decoder is a convolution transpose block which produces an output image $\nnoutput(\gridSet)$ whose dimensions are equal to the input image $\nninput(\gridSet)$. The decision map $\nndecision(\gridSet)$ is obtained as follows
\begin{equation}
	\nndecision(\gridSet) = \mathbbm{1}\left\{\sigma\left(\nnoutput \left(\gridSet \right)\right) >\detThresh \right\},
	\label{eq:nndecisions}
\end{equation}
where $\sigma(.)$ is the sigmoid activation function, and $\detThresh$ is a detection threshold resulting in a binary decision.  The detection threshold $0\leq\detThresh\leq 1$ is generally chosen to be $\theta=0.5$. The number of parameters of the $i$th convolutional layer is $k_i^2 \cdot n_{in,i} \cdot n_{out,i}$, where a $k_i\times k_i$ kernel is used, $n_{in}$ is the number of input feature maps, and $n_{out,i}$ is the number of output feature maps. In addition to the convolutional layers, the batchNorm layers contribute $2n_{in,i}$ parameters at the $i$th layer. The total number of parameters over all layers $L$ is thus $\sum_{i=1}^{L} k_i^2 \cdot n_{in,i} \cdot n_{out,i} + 2n_{in,i}= 29908$.


	
	

\subsection{Dataset}
\label{subsec:dataset}
The training dataset comprises $\nTrain = 20480$ pairs $(\nninput(\gridSet),\nntarget(\gridSet))$ divided evenly according to the number of emitters $\nEmitters \in \left[1,40\right]$, resulting in 512 pairs per $\nEmitters$ per region (for example, San Jose).  The test dataset contains 1024 maps generated in the same way as the training set.  The test and training datasets are generated independently. The emitters are placed randomly (uniformly) over $\region$. The emitter powers are chosen uniformly over the interval $\left[0,2\right]$ Watts. The field $\field(\fieldLoc)$ is then obtained using Atoll, using its internal model \eqref{eq:field}, and the occupancy map with respect to the threshold $\occThresh$ is generated according to \eqref{eq:nntarget}. 

Little is gained by making $\nEmitters$ larger than when full occupancy is achieved.  Hence there is a natural ``upper bound" on $\nEmitters$ that depends on the environment where increasing $\nEmitters$ does not add richness to the occupancy maps.  We find that 40 emitters are enough to cover more than more than $92\%$ of the region in Chicago at $\occThresh = -90\,\rm{dBm}$.  Hence, we generally train with no more than 40 emitters.
The datasets are generated with fixed occupancy threshold $\occThresh$ and $\nSensors$. 

\subsection{Training and performance}

\label{subsec:training}


\begin{table}[!t]
	\caption{Losses and error rates in Chicago, Denver, and San Jose: $\nSensors=100$, $\occThresh=-90\,\rm{dBm}$}
	\label{tab:train-test-stats}
	\centering
	\resizebox{0.49\textwidth}{!}{\begin{tabular}{@{}ccccc@{}} \toprule
			Region & Train $\acc(0.5)$ & Test $\acc(0.5)$ & Train loss & Test loss\\ \midrule
			Chicago & 0.058 & 0.062 & 0.138 & 0.147 \\
			Denver  & 0.139 & 0.148 & 0.315 & 0.336 \\ 
			San Jose & 0.075 & 0.079 & 0.189 & 0.205 \\ \bottomrule
	\end{tabular}}
	\vspace{1ex}
	
	{\raggedright \footnotesize \hspace{1em} Training and testing losses \eqref{eq:ce-loss} and error rates \eqref{eq:acc} after 500 epochs. Observe that Chicago provide the best error rate performance. Note that the emitter locations are not shared between training and testing datasets. \par}
\end{table}

\begin{figure}[!t]
	\centering
	\includegraphics[trim={0.3cm 0.0cm 1.1cm 0.7cm},clip,width = 0.49\textwidth]{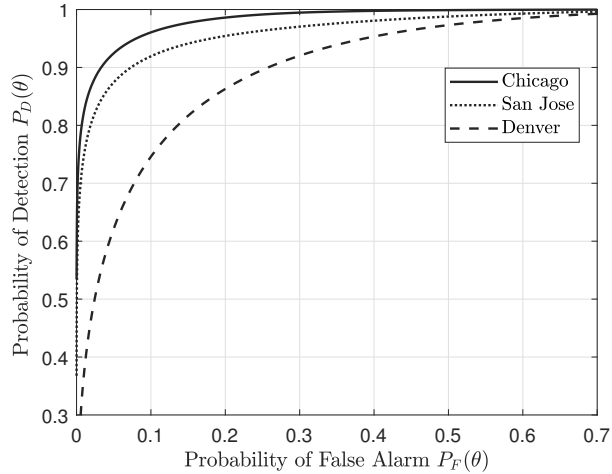}
	\caption{ROC \eqref{eq:pDpF} as $\detThresh$ is varied; in Chicago (solid), Denver (dashed) and San Jose (dotted) with $\occThresh=-90\,\rm{dBm}$. Observe that because its curve is above and to the left of the others, Chicago provides the best ROC performance.}
	\label{fig:roc}
\end{figure}
To supervise the training process, training pairs $(\nninput(\gridSet),\nntarget(\gridSet))$ are sampled randomly from the dataset and grouped into mini-batches of size 32. The corresponding outputs $\nnoutput(\gridSet)$ are obtained from the inputs $\nninput(\gridSet)$ as described in Section \ref{subsec:nn}. During testing, the decision maps $\nndecision(\gridSet)$ are obtained by thresholding as shown in \eqref{eq:nndecisions}. The learning rate is $5\times 10^{-5}$; reduced by a factor of 10 if the loss plateaus over a period of 10 epochs (one complete run through the training set).



We use the weighted cross entropy loss function \cite{aurelio2019learning,jadon2020survey} defined as,
\begin{multline}
	\loss(\posWeight)= \frac{1}{ \nTrain \nGrid} \sum_{j=1}^{\nTrain}\sum_{k=1}^{\nGrid} \Big(- \posWeight \ \nntarget_j(\gridSet_k)\log\left[\sigma(\nnoutput_j(\gridSet_k))\right] 
	- (1 - \nntarget_j(\gridSet_k))\log\left[1-\sigma(\nnoutput_j(\gridSet_k))\right]\Big),
	\label{eq:ce-loss}
\end{multline}
where $\posWeight$ is a factor that weighs the sub-regions $\gridSet_k$ for which $\nntarget(\gridSet_k) = 1$. The weights of the network are tuned by an ADAM optimizer \cite{kingma2014adam} to minimize \eqref{eq:ce-loss}. We train the neural network for 500 epochs after which the neural network weights are saved.

The performance of the decision system during testing is determined by the error rate of classifying the sub-regions $\gridSet_k \in \gridSet$, or
\begin{equation}
	\acc(\detThresh) = \frac{1}{{\nTrain \nGrid}} \sum_{j=1}^{\nTrain} \sum_{k=1}^\nGrid |\nntarget_j(\gridSet_k) - \nndecision_j(\gridSet_k)|.
	\label{eq:acc}
\end{equation}
The expression in \eqref{eq:acc} depends on $\detThresh$ through the decision maps $\nndecision(\gridSet)$, and we set $\detThresh=0.5$.
The error rate is averaged over emitter locations, powers, and numbers, and sensor locations drawn randomly.  The distributions of $\emitterSet$ and $\sensorSet$ are uniform over $\region$.

The training and test error rates $\acc(0.5)$ are shown in Table.~\ref{tab:train-test-stats}. Chicago occupancy maps have lower error rates and loss values than Denver, for example, probably because of the dramatic differences in terrain. San Jose falls in between with the presence of both flat (valley) and irregular topographies. Training and testing are performed on a single \emph{NVIDIA GeForce GTX Titan X} GPU. During testing, the time to produce a decision map $\nndecision(\gridSet)$ from an input $\nninput(\gridSet)$ is approximately $8\,\rm{ms}$.  This value is smaller than typical delays of wireless sensor networks, and thus decisions can be generated in real-time as soon as the measurements are available.

Receiver operating characteristic (ROC) curves, which are plots of detection rates versus false alarm rates, can be generated by varying $\posWeight$ in \eqref{eq:ce-loss} or $\detThresh$ in \eqref{eq:nndecisions}.  We choose $\theta\in(0,1)$, and set $\posWeight=1$, allowing us to avoid retraining the network for every $\posWeight$. The false alarm rate, denoted by $\pF(\detThresh)$, is the rate of incorrectly deciding $\nndecision(\gridSet_k) = 1$, averaged over all $\gridSet$, and the detection rate $\pD(\detThresh)$ is the rate of correctly deciding $\nndecision(\gridSet_k) = 1$, defined as
\begin{align}
	\pF(\detThresh) =\frac{\sum_{k=1}^{\nGrid}\sum_{j=1}^{\nTrain} (1-\nntarget_j(\gridSet_k))\nndecision_j(\gridSet_k)}{\sum_{k=1}^{\nGrid}\sum_{j=1}^{\nTrain}(1-\nntarget_j(\gridSet_k))}, \quad     \pD(\detThresh) = \frac{\sum_{k=1}^{\nGrid}\sum_{j=1}^{\nTrain} \nntarget_j(\gridSet_k)\nndecision_j(\gridSet_k)}{\sum_{k=1}^{\nGrid}\sum_{j=1}^{\nTrain}\nntarget_j(\gridSet_k)}. \label{eq:pDpF}
\end{align}

Fig.~\ref{fig:roc} shows the ROC curves with $\nSensors=100$ and $\occThresh=-90\,\rm{dBm}$ and their dependence on the region; more favorable ROC is achieved in Chicago, than in San Jose or Denver. Even when there is no sensor noise, a nontrivial ROC curve is obtained since our system estimates the occupancy everywhere, including sub-regions in which there are no sensors. 

\begin{figure}[!t]
	\centering
	\includegraphics[trim={0.2cm 0.0cm 1.1cm 0.7cm}, clip, width = 0.49\textwidth]{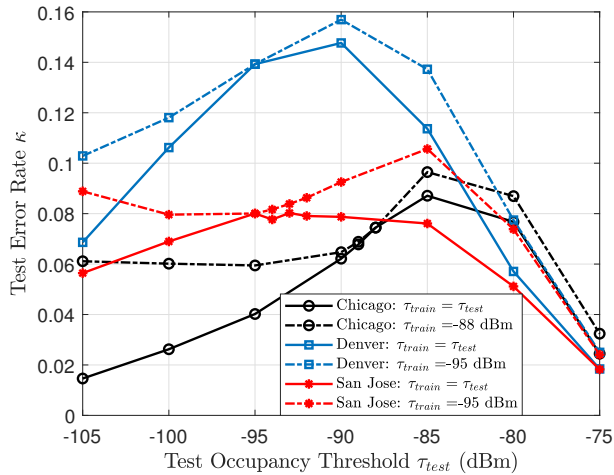}
	\caption{\footnotesize Robustness to $\occThresh$. Solid curves represent matched training and testing: $\occThresh_{train} = \occThresh_{test}  \in [-105,-75] \, \rm{dBm}$; Dash-dotted curves represent mismatched training and testing: fixed $\occThresh_{train}$, but $\occThresh_{test} \in [-105,-75] \, \rm{dBm}$. Robustness is evidenced by observing that the dash-dotted curves are close to the solid ones of the same color over a wide range of $\occThresh_{test}$.  The fixed $\occThresh_{train}$ is chosen to have approximately 50\% of the region occupied.}
	\label{fig:acc-tau-match-mismatch}
\end{figure}

\section{Robustness and Comparison with other Methods} 
\label{sec:robustness}
The usefulness of any neural network-based decision system is determined in part by its robustness to parameter values for which it has not been trained. Until now, we have matched the training and test parameters as we averaged over emitter and sensor realizations. In this section, we examine the test error rate of the decision system when: (1) the occupancy threshold $\occThresh$, (2) the number of sensors $\nSensors$, and (3) the measurement noise have values that differ during testing and training. We show that the decision system is robust to a wide range of $\occThresh$ and $\nSensors$; we can train with a single $\occThresh_{train}$ (or $\nSensors_{train}$), and test with $\occThresh_{test} \ne \occThresh_{train}$ (or $\nSensors_{test} \ne \nSensors_{train}$), and get an error rate similar to when $\occThresh_{train} = \occThresh_{test}$ (or $\nSensors_{train} = \nSensors_{test}$). Moreover, we model noisy power measurements, and show that forming the neural network inputs from LLRs makes the decision system robust to noise, even when trained with noise-free measurements. 

\subsection{Occupancy threshold $\occThresh$}
\label{subsec:occThreshRobust}
The decision maps $\nndecision(\gridSet)$ are always referenced to an occupancy threshold $\occThresh$.  We show that a system trained for a threshold $\occThresh_{train}$ can be used effectively for testing with $\occThresh_{test}\neq\occThresh_{train}$.
In this case, $\sensdecision(\sensorLoc) = \meas(\sensorLoc)-\occThresh_{train}$ during training, and $\sensdecision(\sensorLoc) = \meas(\sensorLoc)-\occThresh_{test}$ during testing. 


The test error rate $\acc$ of the decision system versus $\occThresh_{test}$ are shown in Fig.~\ref{fig:acc-tau-match-mismatch}. For mismatched training and testing, $\occThresh_{train} = -88\,\rm{dBm}$ in Chicago,  $\occThresh_{train} = -95\,\rm{dBm}$ in Denver, and San Jose. Robustness is seen by observing that the dash-dotted curves are close to the solid ones of the same color over a wide range of $\occThresh_{test}$.  

This robustness is attributed to the LLR inputs to the neural network \eqref{eq:sensSoftDec}.  Since $\meas(\sensorLoc) - \occThresh_{test} = \meas(\sensorLoc) - \occThresh_{train} +( \occThresh_{train} - \occThresh_{test})$, the difference $\occThresh_{train} - \occThresh_{test}$ represents the mismatch in measured occupancy between training and testing.  When $\occThresh_{test}<\occThresh_{train}$, this difference represents a simple positive offset to the measurement, which leads to overstating the occupancy. In a similar fashion, $\occThresh_{test} > \occThresh_{train}$ leads to understating the actual occupancy.  

\subsection{Number of sensors $\nSensors$}
\label{subsec:nsensRobust}
The decision system has no control over the number of sensors $\nSensors$ and their locations, and the number of measurements aggregated into $\nninput(\gridSet)$ during training and testing could be different. As described in Section \ref{sec:occ-map}, the sensor locations are varied during the training so that there are no location biases. However, we postulate that, the test error rate of a system trained for $\nSensors_{train}$ will decrease with $\nSensors_{test}>\nSensors_{train}$ because more $\sensdecision(\sensorLoc)$ are aggregated per sub-region. Note that in our settings, $\nSensors$ is sampled from $\region$ which is discretized into $512\times512$ points. 

We follow the same process described in Section \ref{subsec:dataset} for each region, and generate training and test sets for values of $\nSensors_{train}$ and $\nSensors_{test} \in \left[50,800\right]$. We use $\nSensors_{train}=\nSensors_{test}$ sensors for training to establish the baseline performance. Due to the aggregation step \eqref{eq:nninput}, the size of the input $\nninput(\gridSet)$ remains fixed at $128 \times 128$, even as $\nSensors_{test}$ is changed. 

\begin{figure*}[!t]
	\centering
	\subfloat[]{\includegraphics[trim={0.2cm 0.0cm 1.0cm 0.6cm}, clip, width = 0.33\textwidth]{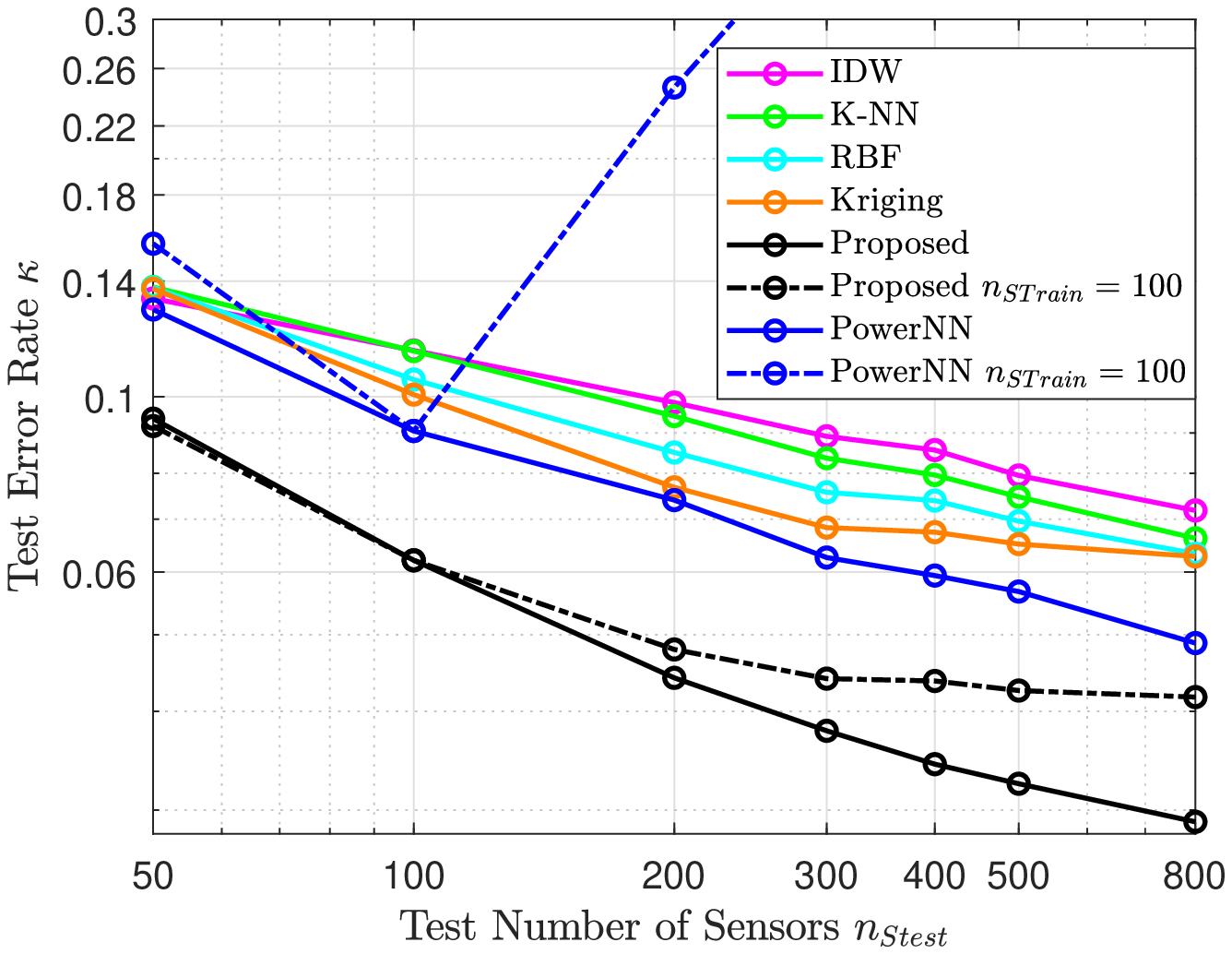}
		\label{fig:acc-ns-chicago}}
	\subfloat[]{\includegraphics[trim={0.2cm 0.0cm 1.0cm 0.6cm}, clip, width = 0.33\textwidth]{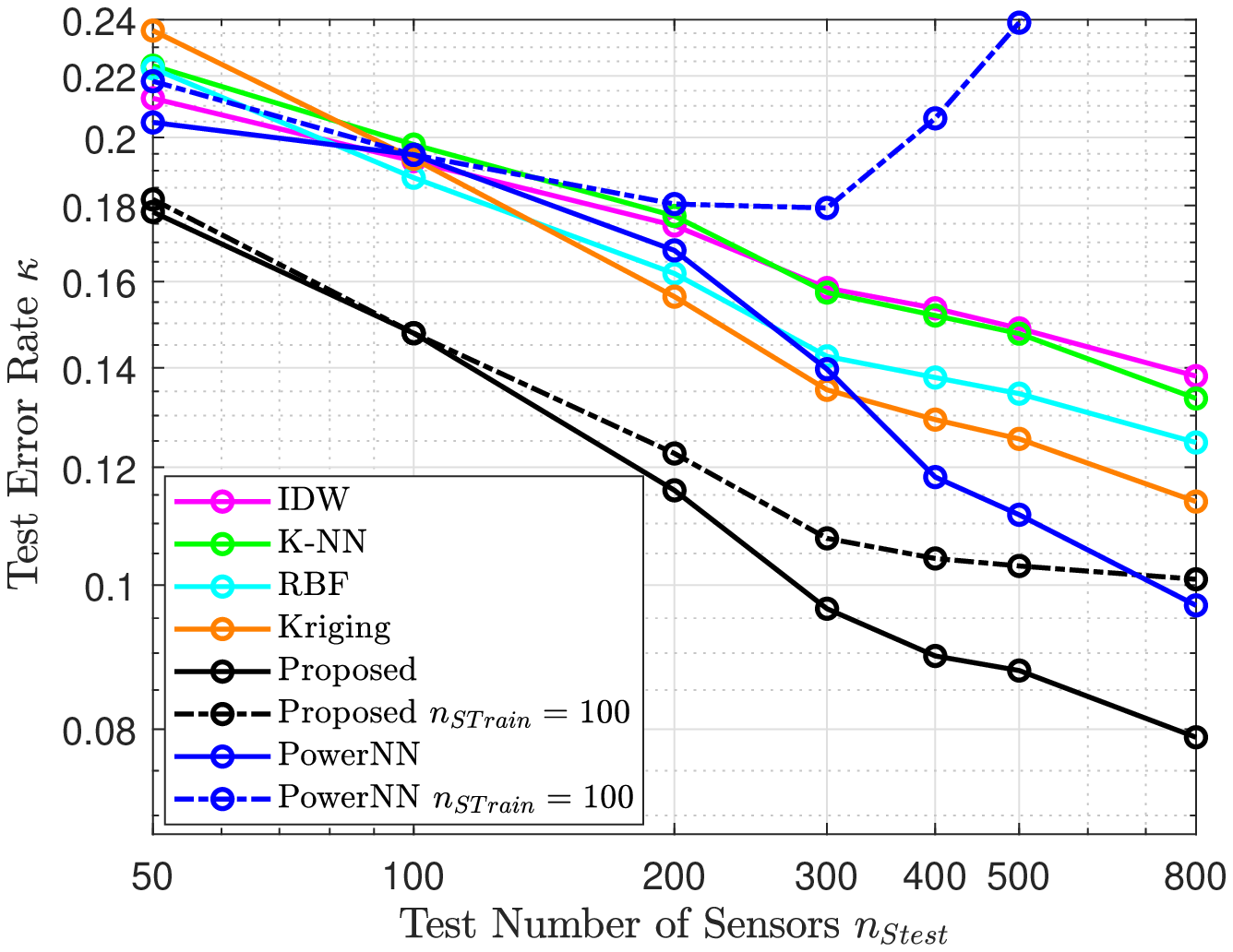}
		\label{fig:acc-ns-denver}}
	\subfloat[]{\includegraphics[trim={0.2cm 0.0cm 1.0cm 0.6cm}, clip, width = 0.33\textwidth]{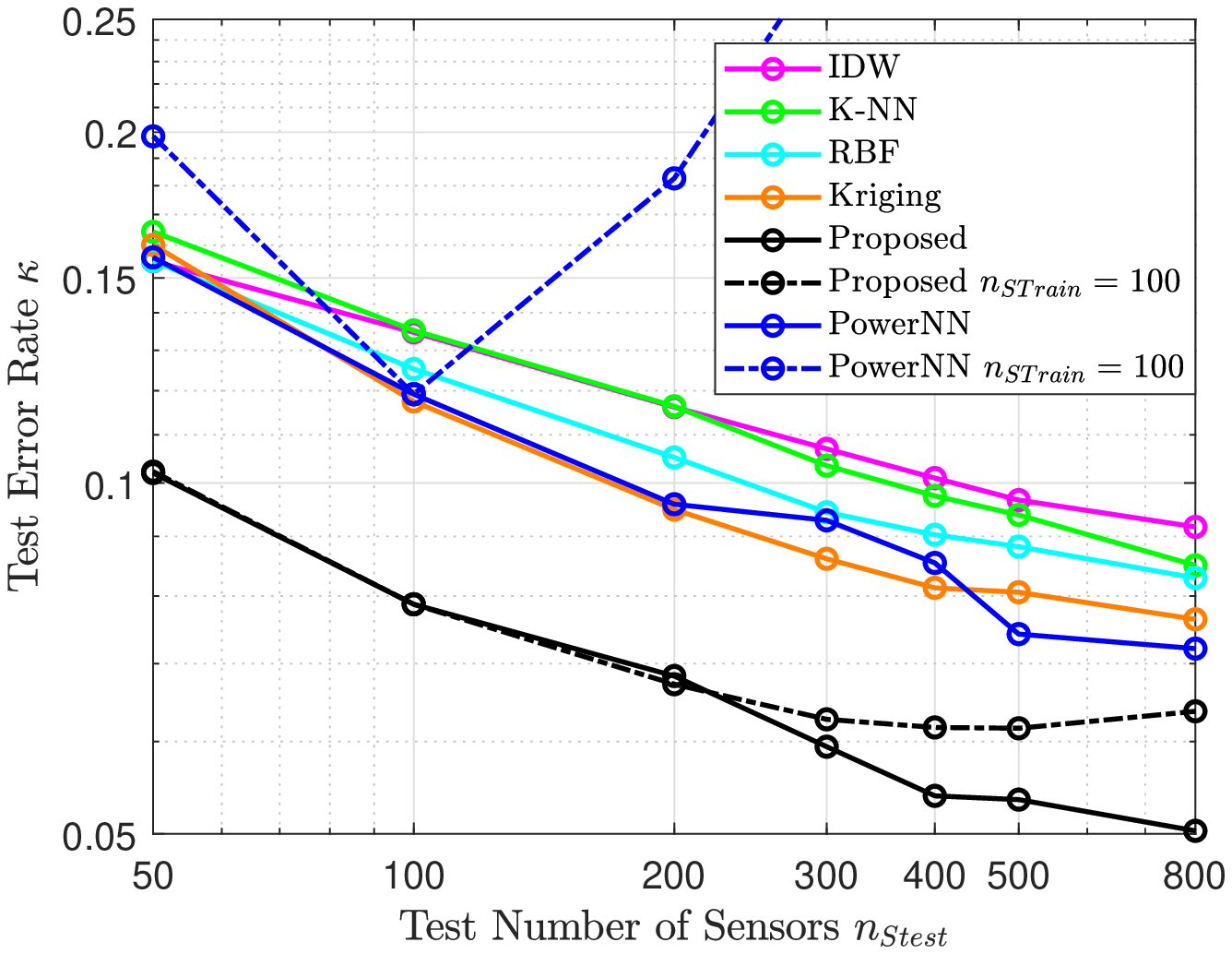}
		\label{fig:acc-ns-sanjose}}
	\caption{\footnotesize Robustness to $\nSensors$. Left: Chicago. Middle: Denver. Right: San Jose. Solid curves represent matched training and testing, $\nSensors_{train}=\nSensors_{test}$. Dash-dotted curves represent mismatched training and testing, $\nSensors_{train}=100$, $\nSensors_{test} \in [50, 800]$.  The proposed method is represented in the black curves. Robustness is seen by observing that the dash-dotted black curves are close to the solid ones for a wide range of $\nSensors$.  Observe from the blue curves that the PowerNN neural network (which is also training-based) is not robust to this mismatch.  The remaining methods do not require training a network.
	}
	\label{fig:acc-ns}
\end{figure*}

To demonstrate the robustness to the number of sensors, we fix $\nSensors_{train} = 100$ during training, and show that increasing the number of sensors during testing is advantageous to the performance, even though there is a mismatch between training and testing.
Fig.~\ref{fig:acc-ns-chicago}, \ref{fig:acc-ns-denver}, and \ref{fig:acc-ns-sanjose} plot the error rate $\acc$ in \eqref{eq:acc} as $\nSensors_{test}$ varies with for each region (black curves).  The solid curves are for matched training and testing, while the dash-dotted curves are for $\nSensors_{train}=100$. Encouragingly, the error rate $\acc$ falls as $\nSensors_{test}$ increases in both cases, and therefore the system is robust. This behavior is not seen with PowerNN, as also shown (blue curves).  This comparison, among others, is discussed in more detail in Section \ref{subsec:compare}. 


\subsection{Noise modeling and robustness}
\label{sec:noise}
Sensor measurements are generally subject to measurement non-linearity and noise. In this section, we analyze the effect of noise on the performance of the decision system.  In particular, we demonstrate that, by modeling measurement noise, the approximate LLRs \eqref{eq:sensSoftDec} can be augmented to improve the performance and make the system robust to mismatched training and testing noise levels.

So far, we have relied on Atoll to simulate the field $\field(\fieldLoc)$. However, Atoll only produces power values, and a model for the received time-varying amplitude is required to model noise. By assuming that the emitters are uncorrelated, the field $\field(\fieldLoc)$ in \eqref{eq:field}, at any location $\fieldLoc \in \region$, can written in the form
\begin{equation}
	\field(\fieldLoc) = \sum_{l=1}^{\nEmitters}\fieldemitter{\emitterLoc_l}{\fieldLoc} \powEl_l,
	\label{eq:fieldlin}
\end{equation}
where $\fieldemitter{\emitterLoc_{l}}{\fieldLoc}>0$ is the power measured by a sensor from a unit-power emitter located at $\emitterLoc_{l} \in \emitterSet$, and $\powEl_l$ is the (unknown) power of the $l$-th emitter (in Watts). We then model the noisy received discrete-time signal at location $\sensorLoc$ as
\begin{equation}
	\recSig(\sensorLoc,\sampleNum) = \sum_{l=1}^{\nEmitters}\sqrt{\fieldemitter{\emitterLoc_l}{\sensorLoc}} \transSig(\sampleNum) + \noise(\sensorLoc, \sampleNum),
	\label{eq:noisySignal}
\end{equation}
where $\sampleNum = 1,2,...,\nSamples$, and $\nSamples$ is the number of samples, $\transSig(\sampleNum)$ are samples of the signal transmitted by the $l$-th emitter modeled as independent complex normal random variables $\mathcal{CN}(0,\powEl_{l})$ for $l = 1,\ 2,\ ...,\ \nEmitters$. 
The sensor thermal noise $\noise(\sensorLoc, \sampleNum) \sim \mathcal{CN}(0,\noisePower)$ are independent and identically distributed random variables for all $\sensorLoc$ and $\sampleNum$, independent of $\transSig(\sampleNum)$ for all $\sampleNum$, and we refer to $\noisePower$ as the noise power.

To estimate the power, the sensors compute the mean-square value of the noisy samples \eqref{eq:noisySignal}, which, assuming the emitters are uncorrelated, and using \eqref{eq:modelNoise}, can be modeled as
\begin{align}
	\meas(\sensorLoc) = \frac{1}{\nSamples}\sum_{n=1}^{\nSamples} |\recSig(\sensorLoc,\sampleNum)|^2 \approx  \meanField + \modelNoise(\sensorLoc) + \noisePower\noise_{sq}(\sensorLoc),
	\label{eq:noisyMeas}
\end{align}
where $2\nSamples \noise_{sq}(\sensorLoc) \sim \chi^2(2\nSamples)$, and $\chi^2(\nSamples)$ denotes a chi-square distribution with $\nSamples$ degrees of freedom. The approximation in \eqref{eq:noisyMeas} assumes that the cross term $\frac{1}{N}\sum_{\sampleNum=1}^{\nSamples}\sum_{l=1}^{\nEmitters} \sqrt{\fieldemitter{\emitterLoc_l}{\sensorLoc}} \transSig(\sampleNum) \noise(\sampleNum)^{*}$ is zero by the weak law of large numbers, where $(\cdot)^{*}$ denotes complex conjugation. Note that as $\nSamples \rightarrow \infty$, $\meas(\sensorLoc) \rightarrow \meanField + \modelNoise(\sensorLoc) + \noisePower$, for $\sensorLoc \in \gridSet_k$, and the error in approximating $\meanField$ by $\meas(\sensorLoc)$ still exists even when the noise is completely averaged out. By the central limit theorem, $\noisePower \noise_{sq} \sim \mathcal{N}(\noisePower,\noiseVar)$, and the sensor measurement $\meas(\sensorLoc) \sim \mathcal{N}(\meanField+\noisePower, \modelVar+\noiseVar)$. 
We assume that the noise power $\noisePower$ is known to the sensor.

\begin{figure}[!t]        
	\centering
	\includegraphics[trim={0.2cm 0.0cm 1cm 0.6cm}, clip, width = 0.49\textwidth]{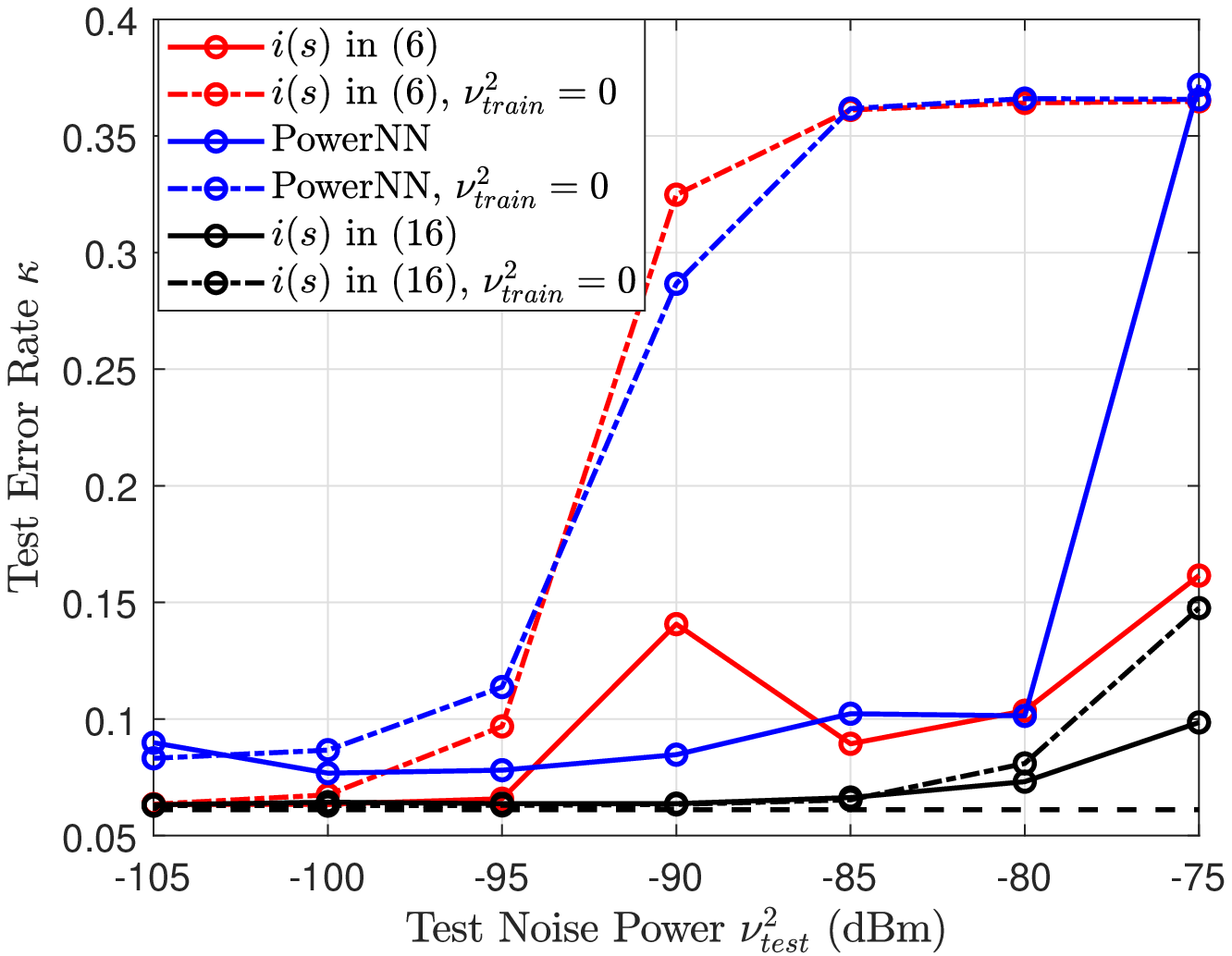}
	\caption{Robustness to $\noisePower$. Solid curves represent matched training and testing: $\noisePower_{train}=\noisePower_{test}$; Dash-dotted curves represent mismatched training and testing: $\noisePower_{train}=0$. Robustness to noise, when \eqref{eq:noisyDecisions} is used for $\sensdecision(\sensorLoc)$, is seen by observing that the dash-dotted black curves are close to the solid ones for a wide range of $\noisePower_{test}$. In contrast, when using \eqref{eq:sensSoftDec} for $\sensdecision(\sensorLoc)$, or with PowerNN, robustness is reduced. The dashed horizontal line is the baseline with noise-free sensors during training and testing, or $\noisePower_{train} = \noisePower_{test} = 0 \,\rm{W}$.}
	\label{fig:acc-noise}
\end{figure}

We take the same approach as in Section \ref{subsection:agg}, and derive the LLRs according to the model \eqref{eq:noisyMeas}. Since $\noiseVar$, and $\modelVar$ are not a function of $\meanField$, the resulting LLRs of the noisy measurements can be expressed as,
\begin{equation}
	LLR(\sensorLoc) = \begin{cases}
		\dfrac{|\meas(\sensorLoc) - \noisePower - \occThresh|(\meas(\sensorLoc) - \noisePower - \occThresh)}{2(\noiseVar + \modelVar)} &  \meas(\sensorLoc) \ge \noisePower,\\
		\hfil \dfrac{\occThresh(2\meas(\sensorLoc) - 2\noisePower - \occThresh)}{2(\noiseVar + \modelVar)} & \text{ otherwise. }
	\end{cases}
	\label{eq:llrKnownVar}
\end{equation}

Similar to \eqref{eq:llrNoiseless}, the denominators in \eqref{eq:llrKnownVar} are constant and can be dropped. Moreover, dropping the absolute value term $|\meas(\sensorLoc) - \noisePower - \occThresh|$ and the constant factor $\occThresh$ from the numerators aids in numerical stability, as measurements generally have many orders of magnitude dynamic range. Thus, the sensors report the approximate LLRs
\begin{equation}
	\sensdecision(\sensorLoc) = \begin{cases}
		\meas(\sensorLoc) - \noisePower - \occThresh & \text{ for } \meas(\sensorLoc) \ge \noisePower,\\
		2\meas(\sensorLoc) - 2\noisePower - \occThresh & \text{ otherwise. }
	\end{cases}
	\label{eq:noisyDecisions}
\end{equation}
The approximate LLRs in \eqref{eq:noisyDecisions} are a generalization of those in \eqref{eq:sensSoftDec}, with equality when $\noisePower=0$.

We demonstrate the performance when using \eqref{eq:noisyDecisions} in Chicago where $\nSensors=100$, $\occThresh=-90\,\rm{dBm}$, and $\nSamples=1024$ for a variety of noise powers $\noisePower$. We denote by $\noisePower_{train}$ and $\noisePower_{test}$ the noise power in the measurements \eqref{eq:noisyMeas} during training and testing respectively. 
We train the decision system by forming the inputs $\nninput(\gridSet$) in \eqref{eq:nninput} from $\sensdecision(\sensorLoc)$ in \eqref{eq:noisyDecisions} with $\noisePower_{train}$ using measurements in \eqref{eq:noisyMeas}, to produce a decision map $\nndecision(\gridSet)$. Testing is done in the same manner, but $\noisePower_{test}$ is used instead.

Similar to the preceding subsections, we consider two cases. In the first case, training and testing are matched, or $\noisePower_{train}=\noisePower_{test} \in [-105,-75]\,\rm{dBm}$, and $\acc$ as a function of $\noisePower_{test}$ is shown in solid black in Fig.~\ref{fig:acc-noise}. In the second case, training and testing are mismatched, or $\noisePower_{train}=0$  and $\noisePower_{test} \in [-105,-75]\,\rm{dBm}$, and the  corresponding $\acc$ is shown in dash-dotted black. Comparing the two curves, we observe that the performance in both cases is similar, and, therefore, the decision system is robust to the mismatch between training and testing noise variance. In fact, the training measurements can be noise-free and the test measurements can be noisy.

The robustness to noise does not hold when noise is not part of the LLR model.  Fig.~\ref{fig:acc-noise} shows the performance (in red) when \eqref{eq:sensSoftDec} is used for $\sensdecision(\sensorLoc)$.  
We observe a divergence starting at $\noisePower_{test}=-95\,\rm{dBm}$ between the red dash-dotted, where training and testing are mismatched, versus the red solid, where training and testing are matched.



\begin{figure}[!t]        
	\centering
	\includegraphics[trim={0.2cm 0.0cm 1cm 0.6cm}, clip, width = 0.49\textwidth]{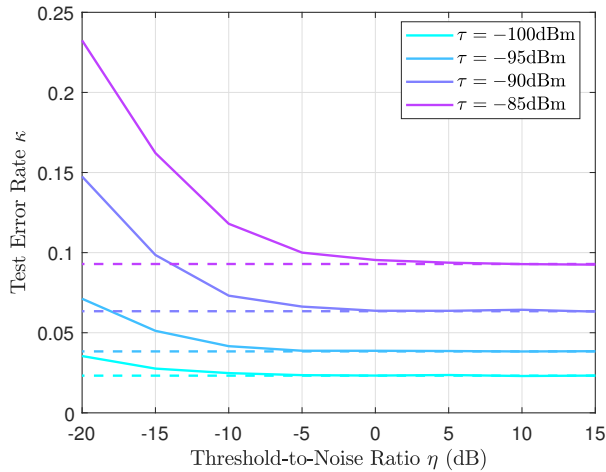}
	\caption{\footnotesize Test error rate versus TNR ($\tnr = \occThresh/\noisePower$ in dB) for various $\occThresh$. Dashed lines indicate error rate with noise-free sensors ($\tnr = \infty$) We can see that low TNR corresponds to high error rate, and the error rate decreases as TNR increases. When $\tnr=0\,\rm{dB}$, the error rates are close to their noise-free limits (dashed lines).}  
\label{fig:acc-tau-tnr}
\end{figure}

\subsection{Comparison with other methods}
\label{subsec:compare}
We compare the error rate of several alternatives for occupancy mapping. We consider \cite{teganya2020deep} as a neural network alternative and we refer to it as PowerNN. PowerNN takes two images, one containing the measurements, which we construct using \eqref{eq:nninput}, replacing $\sensdecision(\fieldLoc)$ with $\meas(\fieldLoc)$ in dBm. Note that for locations with no sensors, $\fieldLoc \in \region \setminus \sensorSet$, $\meas(\fieldLoc)=0$ as per \cite{teganya2020deep}, and we set the ``code length'' to 256. The second image contains zeros for sub-regions $\gridSet_k$ with no sensors, and ones everywhere else. 

We also consider classical alternatives such as inverse distance weighting (IDW) with an exponent of two \cite{shawel2018deep}, and K-nearest neighbors (K-NN) with $K=1$ \cite{cover1967nearest}. We also consider Ordinary Kriging \cite{chakraborty2017specsense}, and radial basis function (RBF) interpolation with a bias term and $\ell_2$-regularization parameter of 0.01 \cite{bishop2006pattern}.

We compare the performance of our proposed method to the alternatives as the number of sensors used in testing is varied $\nSensors_{test}\in[50, \,800]$. We use $\nSensors_{train}=\nSensors_{test}$ sensors for training. The test error rate as $\nSensors_{test}$ varies is shown in Fig.~\ref{fig:acc-ns} for the various methods in all of the considered regions. Our proposed methods outperforms the alternatives for all $\nSensors_{test}$ (solid curves). Note that in Denver, a three-fold increase in the number of sensors is required for the alternatives to perform as well as our proposed method. In Chicago and San Jose, about four-fold increase is required for PowerNN, and at least an eight-fold increase is required for the classical methods.

Moreover, PowerNN is not very robust to the number of sensors or sensor noise. We set $\nSensors_{train} = 100$ during training, and vary the number of sensors during testing $\nSensors_{test}\in[50, \,800]$. Fig.~\ref{fig:acc-ns} plots the error rates of PowerNN for the cases where the number of sensors is matched and mismatched during training and testing. Comparing the solid blue curve to the dash-dotted blue curve, the performance of PowerNN degrades rapidly as $\nSensors_{test}$ increases beyond $\nSensors_{train}=100$. Furthermore, Fig.~\ref{fig:acc-noise} plots the error rates of PowerNN for the cases where the noise is matched and mismatched during training and testing. Comparing the solid blue curve to the dash-dotted blue curve, we observe that under mismatch, with $\noisePower_{test}=-95\,\rm{dBm}$, the error rate increases dramatically compared to the case where noise is matched.

The next section contains some observations on a figure of merit, and comments on practical sensor considerations.

\section{Threshold-to-noise Ratio (TNR) and practical considerations}
\label{sec:tnrOnebit}
\subsection{Threshold-to-noise ratio (TNR) $\tnr = \occThresh/\noisePower$}
\label{sec:tnr}
Classically, the error rate of a noisy detector \eqref{eq:acc} is characterized by the signal-to-noise ratio (SNR), where good performance is generally obtained at high SNR. However, in a spectrum occupancy system, even a low-SNR situation at a sensor (where $\field(\sensorLoc)/\noisePower$ is small) can yield a useful conclusion.  For example, if $\field(\sensorLoc)\ll\noisePower<\occThresh$ we can readily conclude from a small sensor measurement that there is no emitter near the sensor, and hence the sub-region is not occupied.  In this case, a low SNR measurement provides useful information.  On the other hand, if $\noisePower>\occThresh$, then sensor measurements are dominated by noise relative to the threshold, and this noise obscures the presence or absence of an emitter meeting the threshold condition. In this case, a low SNR measurement does not provide useful information.  Hence the SNR of a measurement by itself does not capture the significance of a measurement.



We have found that the threshold-to-noise ratio (TNR), denoted by $\tnr = \occThresh/\noisePower$ is a more meaningful figure of merit to characterize the system performance.  The TNR captures the statistical significance of all the measurements since low TNR values (below 0 dB) indicate that the noise power of the measurements is greater than the threshold, thus giving the measurement reduced significance, irrespective of the signal strength.  Fig.~\ref{fig:acc-tau-tnr} shows the error rate of a system trained and tested with noisy measurements as a function of TNR.  The thresholds are chosen as $\occThresh \in \{-100,-95,-90,-85\}$ dBm, and the noise powers, for every threshold, are chosen such that $\tnr \in [-20,15]\,\rm{dB}$. 
The probabilities of error approaches their limiting value (dashed lines) as $\tnr\rightarrow\infty$, and even with $\tnr=0\,\rm{dB}$ the error rates are close to their limiting values.  Thus, $\tnr$ indicates how close to noise-free performance we are.


\begin{figure}[!t]        
\centering
\includegraphics[trim={0.2cm 0.0cm 1cm 0.6cm}, clip, width = 0.49\textwidth]{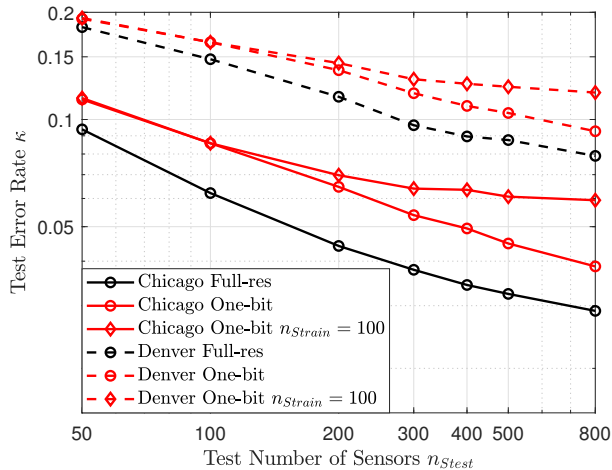}

\caption{\footnotesize Low-resolution (one-bit) sensors work well. Black ($\circ$): full-resolution sensors with $\nSensors_{train} = \nSensors_{test}$. Red ($\circ$): one-bit sensors with $\nSensors_{train} = \nSensors_{test}$.  Red ($\diamond$): one-bit sensors with $\nSensors_{train} = 100$. Comparing the black ($\circ$) with the red ($\circ$) curves shows that only approximately a two-fold increase in the number of one-bit sensors \eqref{eq:sensHardDec} versus full-resolution sensors gives the same $\acc$. Moreover, the red ($\circ$) curves are close to the red ($\diamond$) ones demonstrating robustness to $\nSensors_{test}$. In all cases, $\occThresh=-90\,\rm{dBm}$ and $\noisePower=0$.}
\label{fig:acc-nS-onebit}
\end{figure}

\subsection{Practical sensor considerations and one-bit sensors}
\label{sec:one-bit}
The results of Section \ref{sec:robustness} show that an accurate spectrum occupancy system requires the distribution of tens or hundreds of sensors in a region, which, in turn, requires the sensors to be low-cost and low-power. The results also show that low error rate can be maintained for a wide range of noise power $\noisePower$. In fact, Fig.~\ref{fig:acc-tau-tnr} shows that for any given $\occThresh$, the error rate with noisy sensors is very close to that of noise-free sensors even at TNR as low as $\tnr=-5\,\rm{dB}$. This favorable performance in the presence of noise can be translated into savings on power and cost of the sensors which can be designed with less stringent constraints on gain and noise figure. The RadioHound sensor \cite{kleber2017radiohound,NDRadioHound} developed at the University of Notre Dame is a low-cost spectrum sensor which enables dense deployments. The sensor is in its third generation and features an on-board 8-bit 48 MSps analog-to-digital converter (ADC), with a tunable range of 6 GHz. The sensor has already been used as part of a feasibility study conducted by the Federal Communications Commission (FCC) on the collection of the United States Parcel Service (USPS) broadband data \cite{fccRadiohound}.
The sensor currently consumes about $3.9 \rm{W}$ of power, and costs $\$50$. We plan to reduce the cost and power consumption further by replacing the ADC with a one-bit threshold detector.  In this section, we show that the effect of this modification on performance can be quantified in terms of an increase in number of one-bit versus full-resolution sensors.

Sensors with one-bit threshold detectors report binary (hard) decisions as
\begin{equation}
\sensdecision(\sensorLoc) =
\sgn \left(\meas(\sensorLoc)-\occThresh\right),
\label{eq:sensHardDec}
\end{equation}
instead of real-valued soft decisions \eqref{eq:sensSoftDec}, where the $\sgn(\cdot)$ function returns the sign of its argument and $\sgn(0) = 0$. The hard decisions in \eqref{eq:sensHardDec} do not change the process of aggregation \eqref{eq:nninput}. One intuitively expects the performance of the system to degrade significantly for any $\nSensors$ due to the severe loss of resolution in $\nninput(\gridSet)$.  Fig.~\ref{fig:acc-nS-onebit} compares the performance of full-resolution sensors (black $\circ$) versus one-bit sensors (red $\circ$) for Chicago and Denver. It is remarkable that only approximately a two-fold increase in the number of one-bit versus full-resolution sensors gives the same $\acc$. Moreover, the robustness to $\nSensors_{test}$ with one-bit sensors is observed when comparing the red ($\circ$) curves with the red ($\diamond$) curves.

\section{Conclusion}
\label{sec:conclusion}
We presented a framework for computing occupancy maps for an arbitrary and unlimited number of emitters. We presented a novel log-likelihood ratio approach to input aggregation, which forms fixed resolution images regardless of the number of sensors. We demonstrated that the framework results in accurate maps, and is robust to mismatch between training and testing with respect to the emitter locations, the number of sensors, the occupancy threshold, and noise power. We also defined the threshold-to-noise ratio (TNR) as a figure of merit to capture the performance of a system relative to noise-free sensors. Finally we showed that one-bit sensors, which can provide savings on cost and power, can be used effectively in spectrum mapping.

\section*{Acknowledgements}
The authors thank Vesh Raj Sharma Banjade, Richard Dorrance, and Srikathyayani Srikanteswara of Intel Corporation for helpful discussions leading to this work.

\footnotesize{
\input{main.bbl}

}

\end{document}

%% file: src/header.tex
\usepackage{epsfig,epsf}
\usepackage{color}
\usepackage{subfig}
\usepackage{booktabs,tabularx} 

\usepackage{amssymb}
\usepackage{bbm}
\usepackage{amsfonts}
\usepackage{scalefnt}
\usepackage{amsmath}
\usepackage{bm}
\usepackage{stfloats}
\usepackage{url}

\usepackage{hyphenat}
\usepackage{graphicx}
\usepackage{rotating}
\usepackage[font=footnotesize]{caption}
\usepackage[]{algpseudocode}
\usepackage[ruled]{algorithm2e}
\usepackage[square,numbers,compress]{natbib}

\def\phi{\varphi}

\def\b0{{\mathbf{0}}}

\def\region{\mathcal{R}}                                        

\def\powEl{p}                                                   

\def\meas{m}

\def\sensorEl{s}                                                
\def\sensorLoc{\vect{\sensorEl}}                                   
\def\emitterEl{e}                                               
\def\emitterLoc{\vect{\emitterEl}}                                 
\def\fieldLocEl{r}
\def\fieldLoc{\vect{\fieldLocEl}}                                
\def\field{f}                                                   
       
\def\nninput{I}
\def\sensdecision{i}
\def\nntarget{t}
\def\nnoutput{O}
\def\nndecision{\hat{t}}

\def\denseLayerDepth{\ell}
\def\denseGrowth{\gamma}

\def\acc{\kappa}

\def\sgn{{\operatorname{sgn}}}

\def\transSig{u_{l}}
\def\recSig{y}
\def\sampleNum{n}
\def\noise{w}

\newcommand{\fieldemitter}[2]{\field_{#1}(#2)}

\def\gridSet{\mathcal{G}}                                                  
\def\sensorSet{\mathcal{S}}                                               
\def\emitterSet{\mathcal{E}}                                              
\def\emitterPowersSet{\mathcal{P}}
\def\loss{\mathcal{L}}

\def\posWeight{\alpha}
\def\pD{P_{D}}
\def\pF{P_{F}}

\def\llrNormalize{Z}

\def\nGrid{{n_{\gridSet}}}
\def\nSensors{{n_\sensorSet}}
\def\nEmitters{{n_\emitterSet}}

\def\meanField{\langle \field(\fieldLoc) \rangle_{\gridSet_k}}
\def\modelNoise{z}
\def\modelVar{\zeta^2}

\def\occThresh{\tau}
\def\detThresh{\theta}

\def\nTrain{M}

\def\noisePower{\nu^2}
\def\nSamples{N}
\def\noiseVar{\nu^4/N}

\def\tnr{\eta}
\def\likelihood{q}

\newcommand{\vect}[1]{\mathbf{#1}}